\documentclass{IOS-Book-Article}

\usepackage{mathptmx}
\usepackage{soul}\setuldepth{article}
\def\hb{\hbox to 11.5 cm{}}

\usepackage[T1]{fontenc}
\usepackage{graphicx}
\usepackage{amsfonts} 
\usepackage{stmaryrd}
\usepackage{esvect}
\usepackage{censor}

\usepackage{amsthm}
\theoremstyle{definition}
\newtheorem{definition}{Definition}[section]

\usepackage{tikz}
\usepackage{pgfplots}
\pgfplotsset{compat=1.18}
\usetikzlibrary{shapes.geometric, arrows}
\usetikzlibrary{calc}
\usetikzlibrary{graphs}

\usepackage{xcolor}
\definecolor{codegreen}{rgb}{0,0.6,0}
\definecolor{codegray}{rgb}{0.5,0.5,0.5}
\definecolor{codepurple}{rgb}{0.58,0,0.82}
\definecolor{backcolour}{rgb}{0.95,0.95,0.92}

\usepackage{listings}
\usepackage[nounderscore]{syntax}
\lstdefinestyle{mystyle}{
    backgroundcolor=\color{backcolour},   
    commentstyle=\color{codegreen},
    keywordstyle=\color{magenta},
    numberstyle=\tiny\color{codegray},
    stringstyle=\color{codepurple},
    basicstyle=\ttfamily\footnotesize,
    breakatwhitespace=false,         
    breaklines=true,                 
    captionpos=b,                    
    keepspaces=true,                 
    numbers=left,                    
    numbersep=5pt,                  
    showspaces=false,                
    showstringspaces=false,
    showtabs=false,                  
    tabsize=2
}

\lstdefinelanguage{DSPARQL}{
    morecomment=[l][\color{codegreen}]{\#},
    morestring=[b][\color{blue}]\",
    morekeywords={ SELECT, CONSTRUCT, DESCRIBE, ASK, WHERE, FROM, NAMED, PREFIX, BASE, OPTIONAL, FILTER, GRAPH, LIMIT, OFFSET, SERVICE, UNION, EXISTS, NOT, BINDINGS, MINUS, SIGNALS, WHEN, GROUP, BY, IN, AS, BECOMES, TRUE, FALSE, AT, BIND},
    sensitive=true
}
\lstdefinelanguage{EBNF}{
    morecomment=[l][\color{codegreen}]{\#},
    morestring=[b][\color{blue}]\",
    morekeywords={ ::= },
    sensitive=true
}
\lstdefinelanguage{algebra}{
    morecomment=[l][\color{codegreen}]{\#},
    morestring=[b][\color{blue}]\",
    morekeywords={Extend},
    sensitive=true
}

\usepackage[colorinlistoftodos,prependcaption,textsize=small]{todonotes} 
\usepackage{xargs}    

\newcommandx{\outline}[2][1=]{\todo[inline,linecolor=gray,backgroundcolor=gray!10,bordercolor=gray,#1]{#2{\\\hfill\tiny{Outline}}}}
\newcommandx{\tobias}[2][1=]{\todo[inline,linecolor=cyan,backgroundcolor=cyan!25,bordercolor=cyan,#1]{#2{\\\hfill\tiny{Tobias}}}}

\newcommandx{\todocite}[1]{{\color{red}{[#1]}}}

\newcommand{\domTime}{\mathbb{T}}
\newcommand{\domVal}{\mathbb{D}}

\usepackage[acronym]{glossaries}
\newacronym{ap}{AP}{Active Power}
\newacronym{cep}{CEP}{Complex Event Processing}
\newacronym{cps}{CPS}{Cyber-Physical System}
\newacronym{ebnf}{EBNF}{Extended Backus–Naur Form}
\newacronym{ev}{EV}{Electric Vehicle}
\newacronym{frp}{FRP}{Functional Reactive Programming}
\newacronym{iot}{IoT}{Internet of Things}
\newacronym{iiot}{IIoT}{Industrial Internet of Things}
\newacronym{obda}{OBDA}{Ontology-Based Data Access}
\newacronym{rsp}{RSP}{RDF Stream Processing}
\newacronym{rdf}{RDF}{Resource Description Framework}
\newacronym{soc}{SoC}{State of Charge}
\newacronym{stl}{STL}{Signal Temporal Logic}
\newacronym{tss}{TSS}{Temporal Solution Sequence}

\begin{document}

\pagestyle{headings}
\def\thepage{}
\begin{frontmatter}

\title{SigSPARQL: Signals as a First-Class Citizen When Querying Knowledge Graphs\thanks{This work was conducted within the research project SENSE which was funded by the Austrian Research Promotion Agency FFG (project number 894802).}}

%
%

\markboth{}{July 2025\hb}

\author[A]{\fnms{Tobias} \snm{Schwarzinger}\orcid{0009-0003-1433-2049}%
\thanks{Corresponding Author: Tobias Schwarzinger, \email{tobias.schwarzinger@tuwien.ac.at.}}},
\author[A]{\fnms{Gernot} \snm{Steindl}\orcid{0000-0002-9035-9206}},
\author[A]{\fnms{Thomas} \snm{Frühwirth}\orcid{0000-0001-8133-4747}},
\author[A]{\fnms{Thomas} \snm{Preindl}\orcid{0000-0001-7268-5393}},
\author[B]{\fnms{Konrad} \snm{Diwold}\orcid{0000-0002-6265-4064}},
\author[C]{\fnms{Katrin} \snm{Ehrenmüller}\orcid{0000-0003-1815-8167}},
\author[C]{\fnms{Fajar J.} \snm{Ekaputra}\orcid{0000-0003-4569-2496}},

\address[A]{TU Wien, Vienna, Austria}
\address[B]{Siemens AG \"{O}sterreich, Vienna, Austria}
\address[C]{Vienna University of Economics and Business, Vienna, Austria}

\runningauthor{T. Schwarzinger et al.}

%
\maketitle              
\begin{abstract}
Purpose:
Cyber-Physical Systems (CPSs) integrate computation and physical processes, producing time series data from thousands of sensors.
Knowledge graphs can contextualize these data, yet current approaches that are applicably to monitoring CPS rely on observation-based approaches.
This limits the ability to express computations on sensor data, especially when no assumptions can be made about sampling synchronicity or sampling rates.

Methodology: 
We propose an approach for integrating knowledge graphs with signals that model run-time sensor data as functions from time to data.
To demonstrate this approach, we introduce SigSPARQL, a query language that can combine RDF data and signals.
We assess its technical feasibility with a prototype and demonstrate its use in a typical CPS monitoring use case.

Findings:
Our approach enables queries to combine graph-based knowledge with signals, overcoming some key limits of observation-based methods.
The developed prototype successfully demonstrated feasibility and applicability.

Value:
This work presents a query-based approach for CPS monitoring that integrates knowledge graphs and signals, alleviating problems of observation-based approaches.
By leveraging system knowledge, it enables operators to run a single query across different system instances within the same domain.
Future work will extend SigSPARQL with additional signal functions and evaluate it in large-scale CPS deployments.

\end{abstract}

\end{frontmatter}
\markboth{July 2025\hb}{July 2025\hb}

\section{Introduction}
\label{sec:introduction}

The increase in digitization has led to \glspl{cps}~\cite{baheti2011cyber}, systems that combine computational, communication, and control capabilities with physical components and processes.
Such systems usually encompass a multitude of sensors that are necessary for analysis, monitoring, and control~\cite{liu2017review}.
Examples of \glspl{cps} can be found in domains such as buildings~\cite{akanmu2021towards}, energy networks~\cite{yu2016smart}, and manufacturing~\cite{pivoto2021cyber}.

However, fully utilizing sensor data in \gls{cps} is challenging.
Firstly, data are frequently siloed across various systems.
When this occurs, it is essential to have unified data access to effectively combine information from these sources.
In addition, even if sensor data access does not pose a problem, some tasks require contextualizing the sensor data.
In other words, tasks require information on the \gls{cps} itself, not just the pure sensor data.
For both aspects, Semantic Web technologies can provide a solution.

One solution is \gls{obda}~\cite{poggiLinkingDataOntologies2008}, which enables the creation of a virtual knowledge graph on-top of specialized data stores (e.g., time series databases).
This virtual graph can be queried alongside the context information stored in a graph database, providing contextualized data access without materializing sensor data in the graph.
Such an approach has been applied in \glspl{cps} use cases (e.g., \cite{steindlOntologyBasedOPCUA2019,bakkenChrontextPortableSPARQL2023,kharlamov_semantic_2017}).

Although \gls{obda} solves many issues in \glspl{cps}, computations on time series data can be cumbersome. After mapping a time series to a graph, the temporal values of a single sensor may be spread across thousands of elements that represent individual observations.
This can lead to an increase in the complexity of queries, as the concept of a temporal value must be reconstructed from the individual observations.

While efforts have been made to integrate knowledge graphs and time series data (e.g., \cite{bollen_managing_2024}), they do lack a concept that abstracts over individual observations to alleviate this complexity.
As a result, users must resort to ``workarounds'' to express computations such as summing up the temporal values from two sensors.
Although some requirements can be met with these workarounds, there is a mismatch between the conceptual nature of the computation and the formulation in a query language.
A \emph{Signal}~\cite{hudakArrowsRobotsFunctional2003} is a concept that can abstract over individual observations.

This work proposes to support signals as ``first-class citizens'' when querying knowledge graphs to overcome the shortcomings of contemporary approaches when formulating computations over sensor data.
The integration of this concept involves rethinking the semantics of queries and their results to accommodate the temporal dimension of signals.
When considering this integration, the following research questions emerge.

\textbf{RQ1}. \emph{What are the core operators necessary for integrating knowledge graph query languages with signals?}
To answer this question, we introduce a typical \gls{cps} monitoring use case and identify general requirements for working with signals.
Based on these requirements, we derive the core operators necessary for such an integration.

\textbf{RQ2}. \emph{What is an appropriate extension of the SPARQL query language that integrates these operators in the context of \gls{cps} monitoring?}
We investigate this question by proposing an extension to the syntax and semantics of the SPARQL query language~\cite{sparql11-ql}, incorporating the results from RQ1.
We assess the technical feasibility of this extension by implementing a prototype and apply it to the \gls{cps} monitoring use case from RQ1.

This paper is structured as follows.
Section~\ref{sec:related-work} discusses related work, while Section~\ref{sec:motivating-example} presents a motivating example.
Then, Section~\ref{sec:frp} provides a background on signals, and Section~\ref{sec:modeling-ts-data} discusses their relationship to time series data.
Section~\ref{sec:integrating-signals} presents a list of requirements to bridge the gap between knowledge graphs and signals and proposes a set of operators that address these requirements.
Next, Section~\ref{sec:query-language} applies these concepts to the SPARQL query language.
Section~\ref{sec:implementation} assesses its feasibility with a prototype, while Section~\ref{sec:example-query} applies the proposed SPARQL extension to the motivating example.
Finally, Section~\ref{sec:discussion} reflects on the proposed approach and Section~\ref{sec:conclusion} concludes this work.


\section{Related Work}
\label{sec:related-work}

This section will focus on approaches based on \gls{rdf}, as this work focuses on Semantic Web technologies.
Similar approaches may also be applicable to property graphs.
We have identified four key areas related to our research topics:
(i) RDF Stream Processing,
(ii) Querying temporal data with \gls{obda},
(iii) Combining continuous querying and \gls{obda}, and
(iv) Integrating graphs with time series data. 

\textbf{RDF Stream Processing:}
The main objective of \gls{rsp} is to enable continuous queries on streaming graph data.
This technology has been used for monitoring \glspl{cps} (e.g., \cite{giustozzi_abnormal_2019}).
We can broadly distinguish between two \gls{rsp} approaches: window-based systems and those inspired by \gls{cep}.

On the one hand, window-based approaches (e.g.,~\cite{bollesStreamingSPARQLExtending2008, barbieriCSPARQLSPARQLContinuous2009,le-phuocNativeAdaptiveApproach2011,dellaglioRSPQLSemanticsUnifying2014}), focus on querying streaming data by dividing unbounded streams into bounded relations using window operators.
TEF-SPARQL~\cite{gaoRunningOutBindings2015} extends this paradigm by incorporating facts into the query language.
Using facts, one can remember assertions beyond a single window.
Similar approaches have also been proposed for property graphs~\cite{rost_seraph_2024}.

On the other hand, \gls{cep}-inspired \gls{rsp} systems (e.g.,~\cite{anicicEPSPARQLUnifiedLanguage2011,dao-tranEnrichingCQELSComplex2015}) aim to detect patterns in event streams and aggregate them into high-level events.
These efforts integrate \gls{cep} operations into continuous query frameworks based on SPARQL.
OntoEvent~\cite{maOntoEventOntologyBasedEvent2015} proposes formulating \gls{cep} queries within a knowledge base.
Furthermore, PipeFlow~\cite{salehComplexEventProcessing2015}, proposes embedding the queries in a data flow programming language, while Teymourian et al.~\cite{teymourianSemanticRuleBasedComplex2009} embed SPARQL queries in a logic programming environment.

\gls{rsp} shares many ideas and properties with temporal \gls{rdf} graphs~\cite{gutierrezTemporalRDF2005}.
A temporal \gls{rdf} graph associates a validity interval with each triple.
When querying, this validity interval is considered when evaluating a query at a certain time point.
\gls{rsp} systems also annotate their facts with arrival timestamps but focus on continuously updating their results based on new incoming information which is crucial for monitoring \glspl{cps}.

\textbf{Querying temporal data with \gls{obda}:}
The emergence of the \gls{obda} paradigm, marked by approaches such as Mastro~\cite{calvanese2011mastro} and Ontop~\cite{calvaneseOntopAnsweringSPARQL2017}, has increased the applicability of RDF-based methods in \glspl{cps}.
This success stems from the concept of a virtual knowledge graph layer built on top of existing databases.
Recent \gls{cps}-centered approaches focus on building virtual knowledge graphs for time series data.
Steindl et al.~\cite{steindlOntologyBasedOPCUA2019} demonstrated an approach to integrate time series data from OPC UA, a well-known standard in the manufacturing domain.
In addition, Chrontext~\cite{bakkenChrontextPortableSPARQL2023} proposes an approach for rewriting SPARQL queries to queries on time series databases.
Furthermore, Ontop-temporal~\cite{kalayci2019ontology} extends the querying capabilities with temporal logic formulae.

\textbf{Combining continuous querying and \gls{obda}:}
While some approaches focus exclusively on continuous querying or \gls{obda}, others aim to integrate both paradigms.
Notable are the stream-to-ontology mappings in $\text{SPARQL}_\text{Stream}$~\cite{calbimonteEnablingOntologyBasedAccess2010}, which introduce ontology-based access mechanisms for streaming data.
Furthermore, STARQL~\cite{kharlamov_semantic_2017} provides a comprehensive approach that supports both continuous and historical queries within the context of \gls{obda} based on first-order logic.

\textbf{Approaches integrating graphs with time series data:}
Recently, the combination of graph and time series data has received increasing attention.
Bollen et al.~\cite{bollen_managing_2024} present an approach that extends matching graphs with measurement and time series patterns.
The bindings of these patterns can then be used to access the underlying time series.
Furthermore, Ammar et al.~\cite{ammar_towards_2025} outline a research roadmap with the goal of creating a hybrid data model for graphs and time series data. 

The approach proposed in this work can also be attributed to the latter category.
Our approach differentiates itself from all the mentioned works by relying on signals~\cite{hudakArrowsRobotsFunctional2003} to model temporal values.
Using signals allows us to build on existing work created in the \gls{frp} community and tackle problems that arise from using an observation-based approach (see Section~\ref{sec:modeling-ts-data}).


\section{A Motivating Example}
\label{sec:motivating-example}

\begin{figure}
    \centering
    \includegraphics[width=1\linewidth]{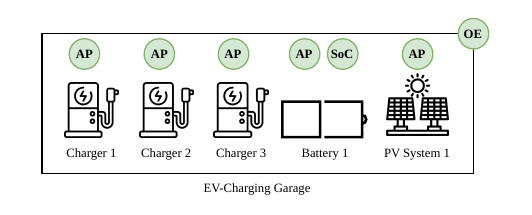}
    \caption{\gls{ev} charging garage with multiple sensors (circles) at different devices (AP~=~Active Power, SoC~=~State of Charge, OE~=~Operating Envelope Limit)}
    \label{fig:use-case}
\end{figure}

This section introduces a motivating example of monitoring a \gls{cps} that is used throughout this work.
Figure~\ref{fig:use-case} shows a conceptual system model of a garage that contains multiple \gls{ev} chargers. 
In addition, the garage includes a photovoltaic system (PV system) and a battery to optimize peak power and overall energy consumption.
However, the garage is a substantial power consumer in the area.
Therefore, without any further restriction, supporting the garage's peak power demands would require additional power grid infrastructure, resulting in additional costs for the involved parties.

An envisioned approach to circumvent additional infrastructure is to impose an operating envelope, limiting the maximum power the garage can drain from the grid.
This limitation places the responsibility of respecting this threshold on the operator of the \gls{ev} chargers, who may be held liable for any violations.
Thus, the garage operator must be able to detect envelope violations to present alerts to the responsible employees.

Detecting envelope violations involves two parts: (i) computing the \gls{ap} of the garage, and (ii) triggering an event once it exceeds the operating envelope.
For the first part, the query must consider observations from all entities within a garage (\gls{ev} chargers, PV systems, and batteries) in the calculation.
Because PV systems are energy producers, the query must invert these sensor readings as they indicate production instead of consumption.
For the latter part, the query must compare the resulting \gls{ap} with the current operating envelope and emit an event for every envelope violation.

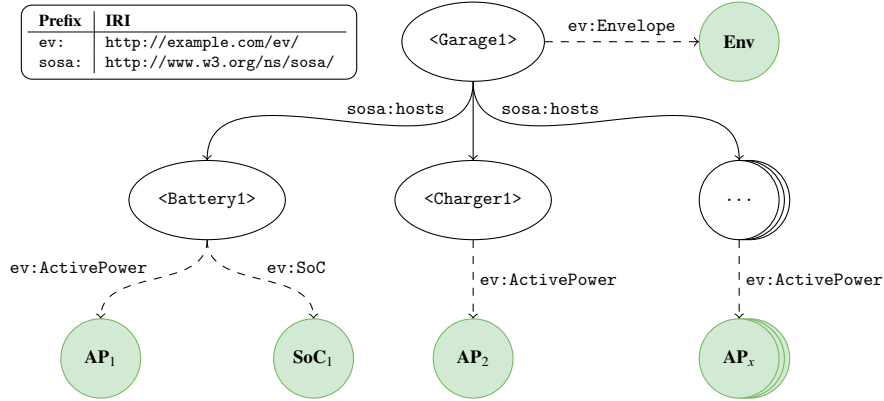
\begin{figure}
    \centering
    \definecolor{lightgreen}{HTML}{d2ead4}
\definecolor{darkgreen}{HTML}{74b75f}

\begin{tikzpicture}[rounded corners, scale=0.7, every node/.style={transform shape}]
\tikzset{
    signal/.style={
        draw=darkgreen,      
        fill=lightgreen,     
        rounded corners=3pt, 
        minimum width=1.5cm,   
        minimum height=1.5cm,  
        font=\bfseries,      
        align=center         
    },
}

\node[draw, anchor=north west, align=left, font=\small] at (-8.5,0.7) {
    \begin{tabular}{l|l}
       \textbf{Prefix}  & \textbf{IRI} \\ \hline
        \texttt{ev:}   & \texttt{http://example.com/ev/} \\
        \texttt{sosa:} & \texttt{http://www.w3.org/ns/sosa/}
    \end{tabular}
};

\node (bg2) at (5.2,-6.0) [ellipse, style=signal] {};
\node (bg1) at (5.1,-6.0) [ellipse, style=signal] {};

\node (bg3) at (5.2,-3.0) [ellipse, draw, minimum width=1.5cm, minimum height=1.5cm, fill=white] {};
\node (bg4) at (5.1,-3.0) [ellipse, draw, minimum width=1.5cm, minimum height=1.5cm, fill=white] {};

\graph[nodes={draw, ellipse, font=\ttfamily, minimum height=1.5cm, minimum width=1.5cm}, edges={font=\ttfamily, in=90, out=270}, no placement]
{
    garage[as=<Garage1>, at={(0,0)}] -> {
        c1[as=<Battery1>, at={(-5,-3)}, >edge label'={sosa:hosts}] -> {
            c1pow[as={$\textbf{AP}_1$}, style=signal, at={(-7,-6)}, >edge label'={ev:ActivePower}, >style=dashed],
            c1plug[as={$\textbf{SoC}_1$}, style=signal, at={(-3,-6)}, >edge label={ev:SoC}, >style=dashed]
        },
        chrg1[as=<Charger1>,  at={(0,-3)}] -> {
            chrg1pow[as={$\textbf{AP}_2$}, style=signal, at={(0,-6)}, >edge label={ev:ActivePower}, >style=dashed]
        },
        others[as={...},  at={(5,-3)}, fill=white, >edge label={sosa:hosts}] -> {
            otherspow[as={$\textbf{AP}_x$}, style=signal, at={(5,-6)}, >edge label={ev:ActivePower}, >style=dashed]
        },
        oe[as={Env}, style=signal, at={(5,0)}, >edge label={ev:Envelope}, >out=0, >in=180, >style=dashed]
    }
};




\end{tikzpicture}
    \caption{Modeling the Garage with the extended data model from Section~\ref{sec:datamodel}. Green nodes represent the signals while the dashed lines represent a mapping from graph nodes and signal properties to the actual signals.}
    \label{fig:datamodel}
\end{figure}

Figure~\ref{fig:datamodel} shows an extended \gls{rdf} graph that models a charging garage.
In addition to a regular \gls{rdf} graph (white ovals and solid lines), the extended data model introduces \emph{signals} (green circles and dashed lines).
For now, it suffices to know that the signals model sensor values and that the dashed connections can associate multiple signals with nodes in the \gls{rdf} graph.
These definitions will be detailed in Section~\ref{sec:datamodel}.

In this example, the \texttt{<Battery1>} has two signals.
$\textbf{AP}_1$ models the active power, while $\textbf{SoC}_1$ models the battery's \gls{soc}.
In addition to the battery, other devices within the garage have an \gls{ap} signal (e.g., \texttt{<Charger1>}). 
Lastly, the garage itself has a signal that models the operating envelope.
As there is no \gls{ap} signal at the garage level, we must compute this signal based on the \gls{ap} of the contained entities (e.g., chargers).
Note that the set of contained devices is not known at design-time and must be extracted by querying the knowledge graph. 
A query that implements this requirement will be presented in Section~\ref{sec:example-query}.

\section{Functional Reactive Programming}
\label{sec:frp}

This section discusses how signals can be formally defined based on existing work in the area of \glsentrylong{frp}~(FRP), such as Fran~\cite{elliott_functional_1997} and Yampa~\cite{hudakArrowsRobotsFunctional2003}.
The central concept in Yampa is the notion of a \emph{signal} -- a value that evolves over time.

Assume that the expression \texttt{SineWave()} in Listing~\ref{lst:example-signals} creates a signal representing a sine wave.
Note that the value of \texttt{a} \emph{implicitly} depends on time.
Therefore, evaluating \texttt{a} at two different time points may yield different results.
As \texttt{a} is a signal, the expression \texttt{b = a * 2} denotes that \texttt{b} is twice the value of \texttt{a} at \emph{any given time point}.
This is different from denoting that \texttt{b} is twice the current value of \texttt{a} in imperative programming languages.
In \gls{frp}, if \texttt{a} changes its value, the signal associated with \texttt{b} automatically reflects this change, allowing \texttt{b} and \texttt{c} to become scaled sine waves.
The concept of a signal is very flexible and allows us to model sensor values over time.

\begin{lstlisting}[
    language=Python,
    caption={An example of working with signals in a simplified language.},
    label=lst:example-signals
]
a = SineWave()
b = a * 2    # a sine wave with bigger amplitude
c = a * 0.5  # a sine wave with smaller amplitude
\end{lstlisting}

Table~\ref{tab:signals} contains descriptions of relevant \gls{frp} concepts and relates these concepts to the motivating example.
Formally, a signal (row 1) is a partial function of a time domain $\domTime$ to a value domain $\domVal$.
The value of a signal $\psi$ at a time point $\tau \in \domTime$ is denoted by $at(\psi, \tau)$.
We refer to any function that has signals as input or output as \emph{signal function}\footnote{This is a more lenient usage than the original definition~\cite{hudakArrowsRobotsFunctional2003}, which restricts signal functions to achieve favorable computational properties. In this work, the objective is to differentiate between regular and signal functions.} (row 2).
Discrete events can be modeled as a sum type that indicates whether an event is present or not (row 3).
The variant of this type representing a present event can carry additional data (e.g., event timestamp).
An event stream denotes a signal of events (row 4).
Lastly, the lift functions allow for lifting scalar values to signals (row 5) and regular functions to signal functions by point-wise application (row 6).

\begin{table}
    \caption{Description of \gls{frp} data types and functions.}
    \label{tab:signals}
    \centering
    \begin{tabular}{|c|l|l|l|}
        \hline
        \textbf{\#} & \textbf{Concept} &  \textbf{Definition}  & \textbf{Example} \\
        \hline
        1 & Signal               & A function from $\domTime$ to $\domVal$ & Active Power over time \\ \hline
        2 & Signal Function      & A function working with signals & Sum of signals  \\ \hline
        3 & Event                & An event is either present or not & \texttt{NoViolation} \texttt{|} \texttt{Violation(e)} \\ \hline
        4 & Event Stream         & A signal of events & Stream of threshold violations \\ \hline
        5 & $\text{Lift}_0(c)$   & Lifts a constant $c$ to a constant signal & Constant -1 to invert producer AP \\ \hline
        6 & $\text{Lift}(f)$     & Lifts a function $f$ to a signal function & $at(a +^\uparrow b, \tau) = at(a, \tau) + at(b, \tau)$ \\ \hline
    \end{tabular}
\end{table}



\section{Modeling Time Series Data as Signals}
\label{sec:modeling-ts-data}

This section discusses problems with observation-based data models and how signals can be used to address these problems.
A time series is usually defined as a sequence of observations $(\tau, d), \tau \in \domTime, d \in \domVal$ with monotonically increasing time stamps.
Although the concept of a time series allows the bundling of all observations of a single sensor, relying on its observation-based nature can cause problems.

\begin{figure}
    \centering
    \begin{minipage}{0.04\textwidth}

\begin{tikzpicture}
    \node[rotate=90] at (0, 0) {Value};
\end{tikzpicture}

\end{minipage}
\begin{minipage}{0.31\textwidth}

\begin{tikzpicture}
\begin{axis}[
    title=a) Linear Interpolation,
    axis lines = left,
    ymin=0, ymax=250,
    xmin=0, xmax=2.25,
    xtick distance=0.5, 
    cycle list name=color,
    scale only axis,
    xlabel={Time},
    width=0.7\textwidth,
]
    \addplot
    coordinates {
        (0,40)(0.5,80)(1,40)(1.5,80)(2.0,40)(2.5,80)
    };
    
    \addplot
    coordinates {
        (-0.25,100)(0.25,100)(0.75,100)(1.25,100)(1.75,100)(2.5,100)
    };
    
    \addplot
    coordinates {
        (0,140)(0.25,160)(0.5,180)(0.75,160)(1,140)(1.25,160)(1.5,180)(1.75,160)(2.0,140)(2.5,180)
    };
\end{axis}
\end{tikzpicture}

\end{minipage}
\begin{minipage}{0.31\textwidth}

\begin{tikzpicture}
\begin{axis}[
    title=b) Last Observation,
    axis lines = left,
    ymin=0, ymax=250,
    xmin=0, xmax=2.25,
    xtick distance=0.5, 
    cycle list name=color,
    scale only axis,
    xlabel={Time},
    width=0.7\textwidth,
]
    \addplot+[const plot]
    coordinates {
        (0,40)(0.5,80)(1,40)(1.5,80)(2.0,40)(2.5,80)
    };
    
    \addplot+[const plot]
    coordinates {
        (-0.25,100)(0.25,100)(0.75,100)(1.25,100)(1.75,100)(2.5,100)
    };
    
    \addplot+[const plot]
    coordinates {
        (0,140)(0.25,140)(0.5,180)(0.75,180)(1,140)(1.25,140)(1.5,180)(1.75,180)(2.0,140)(2.5,180)
    };
\end{axis}
\end{tikzpicture}

\end{minipage}
\begin{minipage}{0.31\textwidth}

\begin{tikzpicture}
\begin{axis}[
    title=c) No Value,
    axis lines = left,
    ymin=0, ymax=250,
    xmin=0, xmax=2.25,
    xtick distance=0.5, 
    cycle list name=color,
    scale only axis,
    xlabel={Time},
    width=0.7\textwidth,
]
    \addplot+[only marks]
    coordinates {
        (0,40)(0.5,80)(1,40)(1.5,80)(2.0,40)(2.5,80)
    };
    
    \addplot+[only marks]
    coordinates {
        (-0.25,100)(0.25,100)(0.75,100)(1.25,100)(1.75,100)(2.5,100)
    };
\end{axis}
\end{tikzpicture}

\end{minipage}
    \caption{Two power sensors (red and blue), and their sum (brown) based on different models for values between observations.}
    \label{fig:time-series-semantics}
\end{figure}
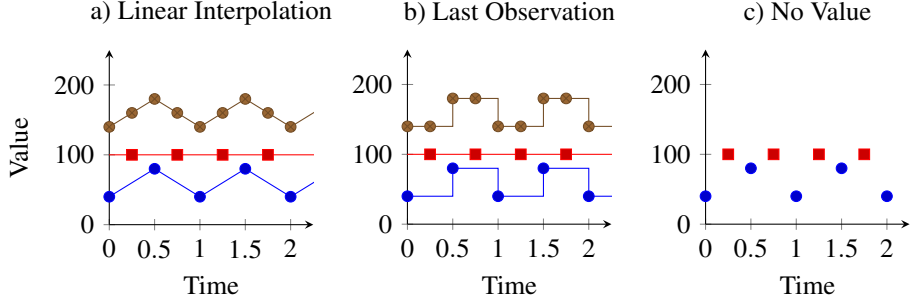

The first challenge arises when combining asynchronous time series, where observations have different timestamps.
Figure~\ref{fig:time-series-semantics} shows the addition of two power sensors and illustrates three ways to model values between observations: linear interpolation (a), last observation (b), and no value (c).
Although all three graphs use the same observations, the observed results differ significantly.

The second problem arises once we consider computations like integration.
These computations require modeling the missing values to obtain meaningful results; otherwise the integral will always be zero since the observations are instantaneous.
Assume that the function $f(\tau)$ models the result of the sum (brown) in Figure~\ref{fig:time-series-semantics} over time $\tau$, the integral $\int_0^{0.5} f(\tau) \, d\tau$ evaluates to $80$~(a), $70$~(b) and $0$~(c), exhibiting different results for each strategy for modeling values between observations.

From an \gls{frp} position, a time series can be modeled as an event stream from $\domTime$ to $\domVal$ that is analogous to (c).
However, based on this event stream, another signal can be created by applying a signal function that models the values between observations.
For example, modeling missing values using the last known observation by ``holding'' the last observation (b) can be expressed using the \texttt{hold} function from~\cite{hudakArrowsRobotsFunctional2003}.
This ability is a powerful tool that allows us to address both of these problems.


\section{A Language-Agnostic Framework}
\label{sec:integrating-signals}

This section introduces a framework of three operators that can be embedded in a knowledge graph query language to support the integration of knowledge graphs and signals.
We first present some requirements and explain their importance.
Then we define operators that fulfill these requirements.
The requirements presented are based on an analysis of the motivating example.
They address the \emph{conceptual core} of such a query language, not particular computational requirements such as computing the mean of a signal.

\begin{itemize}
    \item \textbf{R1} \emph{Creating Signals based on Knowledge Graph Elements}:
    The most essential requirement is that a query language can create signals from knowledge graph elements.
    In the motivating example, this concerns accessing the \gls{ap} of every device in the garage and accessing the operating envelope of a garage.

    \item  \textbf{R2} \emph{Applying Signal Functions}:
    Solely accessing signals does not address many problems when monitoring~\glspl{cps}, as the query must execute further calculations on the sensor data.
    In the motivating example, this requirement concerns computing the total active power of a garage and comparing it to the operating envelope.

    \item \textbf{R3} \emph{Lifting Values to Signals}:
    In many cases, knowledge graph elements or calculated values must be promoted to constant signals, so that they can be used in signal functions.
    In the motivating example, we must compute a factor that can invert the \gls{ap} for power producers.
    This factor must be promoted to a constant signal so that the lifted multiplication operator can be applied.
\end{itemize}
        
Based on these requirements, we introduce a set of operators that form a language-agnostic framework to integrate signals with knowledge graph query languages.
If possible, operators from \gls{frp} are applied to satisfy the requirements.

\begin{itemize}
    \item \textbf{Signal}. The $Signal(s,p)$ operator constructs a signal based on a signal source $s$ and a signal property $p$.
    For example, $Signal(charger1, ActivePower)$ evaluates to the $ActivePower$ signal of $charger1$.
    Language-specific implementations of this operator can restrict the elements of $s$ and $p$ to elements of the used data model.
    For example, Section~\ref{sec:query-language} restricts the elements to IRIs.
    This operator addresses \textbf{R1}.

    \item \textbf{Apply Signal Function}. The $ApplySF(f, \vv{a})$ operator applies an $n$-ary signal function $f$ to a list of signal arguments $\vv{a}$ of length $n$.
    The resulting signal is $f(\vv{a})$.
    This operator can be used for computations on signals.
    This is equivalent to applying a function to the existing signals in \gls{frp}.
    This operator addresses \textbf{R2}.

    \item \textbf{Lift Value}. The $LiftVal(v)$ operator lifts a value $v$ to a constant signal.
    This makes it straightforward to pass elements from the knowledge graph to signal functions.
    This is equivalent to the $\text{Lift}_0$ function in Section~\ref{sec:frp}.
    This operator addresses \textbf{R3}.
\end{itemize}

The list above forms the core operators to bridge the gap between knowledge graphs and signals ($Signal$, $LiftVal$) and to lay the foundation for computations ($ApplySF$).
While the former allows us to obtain signals from the knowledge graph, the latter allows us to make arbitrary computations on the resulting signals, assuming an expressive set of signal functions.
As these two aspects form a conceptual framework for integrating knowledge graph query languages with signals, this operator set answers \textbf{RQ1}.


\section{SigSPARQL: Extending SPARQL with Signals}
\label{sec:query-language}

This section discusses the implications of integrating signals as first-class citizens into an existing knowledge graph query language.
For this purpose, we propose \emph{SigSPARQL}, an extension of the SPARQL 1.1 query language with the operators proposed in Section~\ref{sec:integrating-signals}.
In the remainder of the text, if not mentioned otherwise, we will use the term SPARQL to refer to the SPARQL query language in particular.
This section assumes some familiarity with SPARQL and its algebra as defined in the W3C recommendation~\cite{sparql11-ql}.

Listing~\ref{lst:dsparql-example-select} provides an example.
The query lists all garages with their \gls{ev} chargers.
The \texttt{?ap} variable is bound to a signal that models the \texttt{ev:ActivePower} of the elements bound to \texttt{?charger}.
Binding a variable to a signal models the temporal dimension of the value within a solution.
The remaining section will detail the mentioned concepts.
This example is intended to support the reader by foreshadowing the definitions.

\begin{lstlisting}[
    language=DSPARQL,
    caption={SigSPARQL query for retrieving all garages, their contained chargers, and the \gls{ap} of these chargers. The variables \texttt{?garage} and \texttt{?charger} are bound to RDF terms, while \texttt{?ap} is bound to signals.},
    label=lst:dsparql-example-select
]
SELECT ?garage ?charger ?ap
SIGNALS {
    ev:ActivePower FROM ?charger AS ?ap
}
WHERE {
    ?garage a ev:Garage ;
        sosa:hosts ?charger .
    ?charger a ev:Charger .
}
\end{lstlisting}

\subsection{Data Model}
\label{sec:datamodel}

Before discussing the syntax and semantics of the language, this section formally defines the data model used for evaluating SigSPARQL queries.
Recall Figure~\ref{fig:datamodel} that shows an RDF graph for modeling the \gls{ev} charging garage and its relations to signals.
We restrict ourselves to handling RDF signals as defined in Definition~\ref{def:rdf-sig}.
Evaluating an RDF signal at a given time point will result in an RDF term.

\begin{definition}
\label{def:rdf-sig}
An RDF signal $\psi$ is a (possibly partial) function of the form $\domTime \rightarrow \mathbb{RDF}$ where $\domTime$ is a time domain and $\mathbb{RDF}$ is the set of RDF terms.
\end{definition}

In addition to a standard RDF dataset, the proposed data model contains a set of signals and an annotation function.
While signals model the temporal values in the system, the annotation function associates said signals to elements of the knowledge graph, as depicted in Figure~\ref{fig:datamodel}.
We refer to this triple as \emph{signal-annotated RDF dataset}, as defined in Definition~\ref{def:sg}.

\begin{definition}
\label{def:sg}
A signal-annotated RDF dataset is a triple $<D,S,\varphi>$ where $D$ is a regular RDF dataset, $S$ is a set of RDF signals, and $\varphi$ is a partial function $\mathbb{IRI} \times \mathbb{IRI} \rightarrow S$, where $\mathbb{IRI}$ denotes the set of IRIs.
\end{definition}

The signal annotation function $\varphi$ has two $\mathbb{IRI}$ arguments.
This is necessary because a single IRI can be attributed with multiple signals (e.g., \texttt{<Battery1>} from Figure~\ref{fig:datamodel}).
We refer to the first IRI of $\varphi$ as \emph{signal source} and the second IRI as \emph{signal property}.
In our example, \texttt{<Battery1>} is a signal source while \texttt{ev:ActivePower} and \texttt{ev:SoC} are signal properties.

\subsection{Syntax}
\label{sec:syntax}

SigSPARQL is designed as a superset of SPARQL and adds two clauses: \texttt{SIGNALS} and \texttt{WHEN}.
All grammar rules from the regular SPARQL specification~\cite{sparql11-ql} exist in the grammar of the extension and remain unchanged.
We define the syntax of the additional elements with grammar rules in \gls{ebnf}.

\subsubsection{\texttt{SIGNALS} Clause}

Listing~\ref{lst:grammar-signal} depicts the grammar for the \texttt{SIGNALS} clause.
It consists of a list of signal declarations, each referencing a signal property and a signal source.
The order of the arguments is reversed from the signal annotation function to improve readability.
Finally, each signal declaration contains a variable that binds the resulting signal.

\begin{lstlisting}[
    language=EBNF,
    caption={Syntax of the \texttt{SIGNALS} clause.},
    label=lst:grammar-signal
]
SignalClause ::= "SIGNALS" "{" SignalDeclaration* "}"
SignalDeclaration ::= Iri "FROM" Var "AS" Var
\end{lstlisting}

\subsubsection{\texttt{WHEN} Clause}

Listing~\ref{lst:grammar-when} depicts the grammar for the \texttt{WHEN} clause.
This clause is an expression that evaluates to a Boolean signal.
If the result of the expression is not Boolean, an error is raised.
We refer to this expression as \emph{predicate}.
The expression defines the conditions that trigger the substitution of the graph template when running a continuous \texttt{CONSTRUCT} query (see Section~\ref{sec:semantics}).
The time of these substitutions can be bound to a variable so that it is available in the \texttt{CONSTRUCT} template as a variable.

\begin{lstlisting}[
    language=EBNF,
    caption={Syntax of the \texttt{WHEN} clause.},
    label=lst:grammar-when
]
WhenClause ::= "WHEN" "{" Expression BecomesTrue? "}"
BecomesTrue ::= "BECOMES" "TRUE" ("AT" Var)?
\end{lstlisting}

The \texttt{SIGNALS} clause can only be used in the context of primary \texttt{SELECT} and \texttt{CONSTRUCT} queries, whereas the \texttt{WHEN} clause can only be used in \texttt{CONSTRUCT} queries.
Therefore, the syntax of \texttt{ASK} and \texttt{DESCRIBE} queries is equivalent to regular SPARQL.
A complete grammar of SigSPARQL can be found in the prototype discussed in Section~\ref{sec:implementation}.

\subsection{Semantics}
\label{sec:semantics}

This section describes the semantics of the proposed extension for \emph{static} knowledge bases.
The central idea of the proposed extension is to allow solutions to bind variables to either an \gls{rdf} term or an \gls{rdf} signal.
For example, in Listing~\ref{lst:dsparql-example-select}, the variable \texttt{?charger} binds to the IRIs that identify \gls{ev} chargers, while the variable \texttt{?ap} binds to a signal that models their active power consumption over time.
The semantics of the query language is based on two pillars: the regular SPARQL algebra and \gls{frp}.
The regular \gls{rdf} dataset, which is part of the signal-annotated \gls{rdf} dataset, is utilized for the evaluation of regular SPARQL constructs.
Consequently, SigSPARQL does not modify the semantics of graph pattern matching and filtering solution sequences.

\subsubsection{Temporal Solution Sequences}

Based on the idea of binding variables to signals, we introduce \emph{\glspl{tss}} -- the result of running a \texttt{SELECT} query with a \texttt{SIGNALS} clause.
In a \gls{tss}, solutions bind variables to an \gls{rdf} term or an \gls{rdf} signal.
A \gls{tss} can be evaluated at time point $\tau$, by evaluating $at(\psi, \tau)$ for all bound signals $\psi$.
The resulting solution sequence does not contain any signals, as the evaluation of \gls{rdf} signals results in \gls{rdf} terms.
If $\psi$ is not defined for $\tau$, the variable remains unbound.

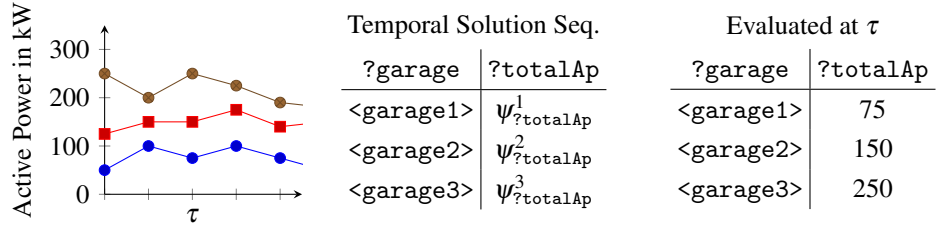
\begin{figure}
    \centering
    \begin{minipage}{0.30\textwidth}
\centering
\begin{tikzpicture}
\begin{axis}[
    axis lines = left,
    ylabel = {Active Power in kW},
    ymin=0, ymax=350,
    xmin=0, xmax=2.25,
    xtick distance=0.5, 
    xticklabels={,,,$\tau$,},
    cycle list name=color,
    scale only axis,
    width=0.7\textwidth,
]
    \addplot
    coordinates {
        (0,50)(0.5,100)(1,75)(1.5,100)(2.0,75)(2.5,50)
    };
    
    \addplot
    coordinates {
        (0,125)(0.5,150)(1,150)(1.5,175)(2.0,140)(2.5,150)
    };

    \addplot
    coordinates {
        (0,250)(0.5,200)(1,250)(1.5,225)(2.0,190)(2.5,180)
    };
    
\end{axis}
\end{tikzpicture}
\end{minipage}
\hfill
\begin{minipage}{0.30\textwidth}
    \centering
    
    Temporal Solution Seq.
    
    \vspace{0.1cm}

    \setlength{\tabcolsep}{0.25em} 
    \renewcommand{\arraystretch}{1.25}
    \begin{tabular}{c|c}
         \texttt{?garage}         & \texttt{?totalAp} \\ \hline
         \texttt{<garage1>}       &  $\psi_\texttt{?totalAp}^{1}$ \\
         \texttt{<garage2>}       &  $\psi_\texttt{?totalAp}^{2}$ \\
         \texttt{<garage3>}       &  $\psi_\texttt{?totalAp}^{3}$
    \end{tabular}
\end{minipage}
\hfill
\begin{minipage}{0.30\textwidth}
    \centering
    
    Evaluated at $\tau$
    
    \vspace{0.16cm}
    
    \setlength{\tabcolsep}{0.25em} 
    \renewcommand{\arraystretch}{1.25}
    \begin{tabular}{c|c}
         \texttt{?garage}   & \texttt{?totalAp} \\ \hline
         \texttt{<garage1>} &  $75$ \\
         \texttt{<garage2>} &  $150$ \\
         \texttt{<garage3>} &  $250$
    \end{tabular}
\end{minipage}

    \caption{Illustrates a \gls{tss} that compute the active power for three garages:  \texttt{<garage1>} (blue), \texttt{<garage2>} (red), \texttt{<garage3>} (brown). The graph shows the temporal evolution of the resulting signals, the center table depicts the \gls{tss} with bound signals, and the right table shows the result of evaluating the \gls{tss} at $\tau$.}
    \label{fig:temporal-solutions}
\end{figure}

Figure~\ref{fig:temporal-solutions} illustrates the concept of \glspl{tss} based on computing the active power of a garage in the motivating example.
The graph in the upper left corner shows the temporal evolution of the \gls{ap} signals for three garages.
The markers on the lines of the figure denote the data points of the underlying signal and are shown for illustration purposes.
The \texttt{?garage} variable is directly bound to an \gls{rdf} term.
In contrast, all solutions bind the \texttt{?totalAp} variable to the signal that represents the computations of the query.

\subsubsection{\texttt{SIGNALS} and \texttt{WHEN} Clause}

The \texttt{SIGNALS} clause is the only means for users to directly create a signal.
All signal declarations result in the algebraic expression $Signal(s,p)$, where $s$ is the signal source, $p$ is the signal property, and $Signal$ is a new expression that implements the operator from Section~\ref{sec:integrating-signals}.
For example, $Signal(\texttt{?garage}, \texttt{ev:ActivePower})$ obtains the \gls{ap} signal of the term bound to \texttt{?garage}.
Evaluating $Signal(s,p)$ for a solution will result in an error if $s$ is unbound, $s$ is not an IRI, or if $Signal(s,p)$ is undefined.

The $Signal$ expression is embedded in an $Extend$ graph pattern from the SPARQL algebra~\cite{sparql11-ql}.
$Extend$ can be used to evaluate an expression and bind the result to a new variable for every solution.
For example, the SPARQL construct \texttt{BIND(?a * ?b AS ?c)} computes the value of \texttt{?a * ?b} and binds it to the variable \texttt{?c}.
The resulting $Extend$ graph patterns are then inserted right before or immediately after the grouping of solution sequences, depending on whether the signal source is a group condition.

For example, in the motivating example, one must sum up all active power sensors in the garage and compare it to the current operating envelope of the garage.
This requires grouping the solution sequence by adding a group condition (e.g., \texttt{GROUP BY ?garage}).
In this scenario, the $Extend$ graph pattern that uses the \gls{ap} sensors (not a grouping condition) is inserted before the grouping, while the $Extend$ graph pattern that uses \texttt{?garage} is inserted after the grouping.
Note that none of these variables are within the scope of the \texttt{GROUP BY} and \texttt{HAVING} clauses, even if they are inserted before the grouping.

In addition to the \texttt{SIGNALS} clause, a \texttt{WHEN} clause can be integrated into the SPARQL algebra by creating an $Extend$ graph pattern that wraps the entire query.
This $Extend$ therefore runs at the end of the query evaluation process and applies the \texttt{WHEN} clause's predicate to every solution.
Based on the resulting signal, an event stream is created that contains an event when the predicate signal changes from false to true (positive edge).
In the motivating example, the positive edge denotes every time point where the envelope \emph{becomes} violated.
The payload of the events is a regular SPARQL solution that is obtained by evaluating the \gls{tss} at the positive edge.
If there is a \texttt{BECOMES} \texttt{TRUE} \texttt{AT} statement in the \texttt{WHEN} clause, an additional binding with time information is added to the solution.
We will use the term \emph{trigger event} to refer to the resulting events.
See the discussion on the \texttt{predicate} function in Fran~\cite{elliott_functional_1997} for a detailed discussion of this concept.

\subsubsection{Signal Computations}

In order to facilitate computations over signals, the SPARQL algebra was further extended with expressions that represent the $LiftVal$ and $ApplySF$ operators from Section~\ref{sec:integrating-signals}.
The proposed extension contains two versions of existing SPARQL functions and operators: one defined for \gls{rdf} terms and a lifted one defined for \gls{rdf} signals.
The regular SPARQL functions and operators are lifted via point-wise application, as shown in Section~\ref{sec:frp}.
The evaluation of the lifted version $f$ is defined via $ApplySF(f, \vv{a})$.
Query engines must automatically use the lifted SPARQL functions and operators if any of their arguments refer to a signal.
As the only way to introduce signals is the \texttt{SIGNALS} clause, query engines can obtain the necessary information by looking at the signal declarations and tracking the usage of the variables.

Whenever a lifted SPARQL operator or function is used, all of its non-signal arguments must be lifted by using a $LiftVal$ expression.
This ensures that, for a given function or operator, all arguments are either \gls{rdf} terms or \gls{rdf} signals.
The $Extend$ graph patterns are inserted according to the regular SPARQL behavior.

\subsubsection{An Example of Evaluating an Expression}

Assume that a query contains the expression \texttt{?ap * ?sign}, where \texttt{?ap} is a signal referring to the \texttt{ev:ActivePower} of a \texttt{?device} and \texttt{?sign} refers to an \gls{rdf} term indicating whether \gls{ap} must be inverted (for PV systems).
Furthermore, assume that the result of the expression is bound to the \texttt{?real\_ap} variable.
As one subexpression of \texttt{?ap * ?sign} is a signal, the lifted version of \texttt{*} is used.
Then, since \texttt{?sign} is a regular \gls{rdf} term, the \texttt{?sign} variable is wrapped in a $LiftVal$ expression to ensure that all arguments are signals.
Listing~\ref{lst:algebra} depicts the nested $Extend$ graph patterns for the expression.

\begin{lstlisting}[
    language={algebra},
    caption={Representation of \texttt{?ap * ?sign} in the extended SPARQL algebra, where \texttt{?ap} is a signal that refers to the \texttt{ev:ActivePower} of \texttt{?device} and \texttt{?sign} is a lifted value.},
    label=lst:algebra
]
# [...] Outer Query
Extend:
    variable: ?real_ap
    expression: ApplySF(*, ?ap, ?sign)
    Extend:
        variable: ?sign
        expression: LiftVal(?temp_sign) # ?temp_sign is 1 or -1
        Extend:
            variable: ?ap
            expression: Signal(?device, ev:ActivePower)
            # [...] Inner Query (e.g., Pattern Matching)
\end{lstlisting}

\subsubsection{Query Forms}

SigSPARQL queries that have a \texttt{SIGNALS} clause enable continuous queries.
The result of a continuous \texttt{SELECT} query is a temporal result set equal to the \gls{tss} of the outermost graph pattern of the query.
This result set also includes bindings to the same signals.
As the signals model temporal values, users can evaluate the signals in the result set at different time points analogously to evaluating a \glspl{tss}.
The result of a \texttt{CONSTRUCT} query is the set union of all triples by substituting the template with the solution variables in the trigger events.
In other words, the result is a continuously growing \gls{rdf} graph, as a single solution can cause multiple trigger events.

In regular SPARQL, blank nodes in the template are scoped to every solution~\cite{sparql11-ql}.
In continuous \texttt{CONSTRUCT} queries, blank nodes in the template are scoped to each trigger event.
Therefore, blank nodes are scoped to each template substitution, similarly as in regular SPARQL.
This allows users, for example, to use blank nodes to describe operating envelope violations.
As each violation has a distinct trigger event, the resulting blank nodes will be different due to the scoping rules.
The query in Section~\ref{sec:example-query} uses the adapted scoping rules to generate a distinct blank node for each violation.

\section{Feasibility Evaluation}
\label{sec:implementation}

We have implemented the proposed SPARQL extension in a prototype\footnote{\url{https://doi.org/10.5281/zenodo.15260651}} based on Oxigraph~\cite{pellissiertanonOxigraph2023} to demonstrate its technical feasibility.
The implementation uses discrete time and models the data between observations using the ``last observation'' strategy.
The evaluation of continuous queries is done in two steps.

Firstly, the query evaluator constructs a description of how to compute the bound signals.
For example, encoding the expression $Signal(\texttt{<Charger1>}, \texttt{ev:ActivePower})$ in the result set, the system encodes the intention that the signal property \texttt{ev:ActivePower} should be obtained from \texttt{<Charger1>}.
These expressions do not refer to variables but to concrete IRIs.
This step stores the information needed to retrieve the bound signals.

Then, this result can be registered in a continuous query engine.
The continuous query engine can maintain the intermediate result of the query (i.e., the result of evaluating regular SPARQL constructs), including the $Signal$ expressions mentioned above.
The continuous query engine then registers for updates on the underlying signals and processes incoming updates.
The prototype facilitates this by inserting newly arriving values into a time series that holds the underlying data for the signal.
Based on this time series, the signal models the values between the data points.
These updates also trigger the evaluation of the $ApplySF$ operators in the query.

After a query has been registered, the temporal result set for \texttt{SELECT} queries can be evaluated at different time points, while the graph result of \texttt{CONSTRUCT} queries is automatically updated for every detected trigger event.
The implementation contains a test suite with multiple queries in the context of the motivating example.


\section{Proof of Concept}
\label{sec:example-query}

In this section, we demonstrate that the proposed query language, and thus the underlying language-agnostic framework, can be used to address the motivating example.
Recall that we want to construct a query that continuously computes the active power consumption for each garage and detects once their power consumption exceeds their operating envelope.
Listing~\ref{lst:dsparql-example} depicts such a query.
We discuss the query in the order of execution in the extended SPARQL algebra.
The \texttt{WHERE} clause (lines 15-19) is used to discover the garages and their contained devices.
For every device, the query evaluates whether the device is a photovoltaic system and binds $1$ or $-1$ to \texttt{?sign} (line 18).

\begin{lstlisting}[
    language=DSPARQL,
    caption={Query to detect events that indicate that a garage exceeded the maximum allowed amount of power consumption. Prefix declaration omitted for brevity.},
    label=lst:dsparql-example
]
CONSTRUCT {
    ?garage ev:hasEnvelopeViolation [
            ev:description "Envelope Violated!" ;
            ev:startTime ?violation_time
        ]
}
WHEN {
    SUM(?ap * ?sign) > ?env
    BECOMES TRUE AT ?violation_time
}
SIGNALS {
    ev:ActivePower FROM ?device AS ?ap
    ev:Envelope FROM ?garage AS ?env
}
WHERE {
    ?garage a ev:Garage ; sosa:hosts ?device .
    ?device a ?ap_device_type .
    BIND(IF(?ap_device_type = ev:PVSystem, -1, 1) AS ?sign)
}
GROUP BY ?garage
\end{lstlisting}

The query then obtains the \texttt{ev:ActivePower} signal for each \texttt{?device} and binds the resulting signal to the \texttt{?ap} variable (line 12).
Then, all solutions (including the active power signals) are grouped according to the garage that contains them (line 20).
After grouping, the signal that describes the \texttt{ev:Envelope} of the \texttt{?garage} is bound to the \texttt{?env} variable (line 12).
The condition for firing trigger events is determined by the \texttt{WHEN} clause (lines 7-10), while the \texttt{CONSTRUCT} clause (lines 1-6) defines the triple template.
The time point of each trigger event is bound to \texttt{?violation\_time} (line 9).
This information can be used as the time stamp of the violation.
While this query is running, the result is a continuously growing \gls{rdf} graph that describes all envelope violations.
Note that due to the scoping rules for blank nodes in the \texttt{CONSTRUCT} clause, each violation (trigger event) becomes a distinct node in the graph.
In combination with Section~\ref{sec:implementation}, we have provided an answer to \textbf{RQ2} by demonstrating that it is feasible to implement the proposed SPARQL extension and that it can address a typical \gls{cps} monitoring use case.


\section{Discussion}
\label{sec:discussion}

The ability to use knowledge graphs to define continuous queries via signals has potential in monitoring \gls{cps}. 
This is because often two \glspl{cps} are slightly different from each other, even though they are in the same domain. 
For example, in the motivating example, an operator may own hundreds of garages.
Although they will adhere to the same rules, many garages will be different (e.g., number of chargers).
The proposed approach allows operators to formulate a single query that can be applied to all their charging garages.

Unfortunately, supporting all possible types of signals can quickly become complex to implement.
For example, when supporting ``linear interpolation'' as a way to model the values between observations, operations can become more complex.
Then, checking inequalities such as $x \geq 0$ in a \texttt{WHEN} clause must also consider these semantics and pinpoint the exact time point where $x$ crosses zero.
One way to circumvent this problem is to only support a restricted set of signals that allow for an efficient query evaluation (like ``last observation'' in our prototype).
In addition, given a function to sample signals, users can fall back to observation-based approaches when needed.
Furthermore, one can build on work from the \gls{frp} community to address the increased complexity and some formalisms for monitoring \glspl{cps} already assume piecewise linear functions (e.g., \cite{maler_monitoring_2004}).

In addition, the proposed extension is not expressive enough to specify computations, such as ``the average of this signal in the last 10 minutes.''
This limited expressiveness is solely due to the fact that this work does not propose new signal functions beyond the point-wise SPARQL built-ins.
Hallé~\cite{halle_complex_2017} argued that a language for event processing should only provide ``syntactic glue'' to combine different event processing languages.
The reason for this is that it is not possible to define a single specification language that ergonomically addresses all the requirements of all possible use cases in all possible domains.
This is also true for monitoring \gls{cps}, as one can deduce from the variety of (vastly different) approaches from the runtime verification community (e.g., \cite{maler_monitoring_2004,brimSTLExtendingSignal2014a,dangeloLOLARuntimeMonitoring2005}) without even considering approaches from other communities (e.g., \gls{cep}).
Although this work does not present a system that enables multiple event detection methods, by relying on signals that can model continuous and discrete temporal values, we propose an approach that can become such an envisioned glue language.

Finally, this work does not include a performance analysis of the prototype, as the focus is on the query language itself, and performance can be significantly influenced by the maturity of the implementation (i.e., degree of optimization).
Future research is necessary to explore optimized implementations of the proposed approach.
Especially, when considering the support of event detection methods for monitoring \gls{cps} that already assume piecewise linear functions, we see no reason why the proposed approach cannot achieve competitive performance.


\section{Conclusion \& Future Work}
\label{sec:conclusion}

This work proposes an approach for querying knowledge graphs and signals.
While the knowledge graph captures information on the \gls{cps}, signals provide a powerful means of modeling run-time sensor data.
We have demonstrated the feasibility and applicability of this approach by defining, implementing, and applying the query language SigSPARQL.

The proposed approach enables context-aware monitoring of \glspl{cps}, allowing users to pose the same queries across multiple system instances, as knowledge about the concrete system can be accessed within a query.
Real-world adoption of such an approach requires further research and development, including a mature and optimized prototype and adaptations of the surrounding technology stack.
For example, the SPARQL protocol must be extended to support interactions with continuously running queries and asynchronous updates.
These efforts may be justified by the potential benefits in energy efficiency and sustainability that comprehensive monitoring can bring to \glspl{cps}.

Concrete possible future directions include (i) designing an optimized prototype for performance evaluations and large-scale monitoring use cases, (ii) adding advanced signal processing functions (e.g., integration), (iii) conducting user evaluations to assess usability benefits of the language, (iv) extending the query language's semantics to temporal knowledge graphs, and (v) investigating the requirements of real-world \gls{cps} deployments and how such an approach can benefit them.


%
%
%
\bibliographystyle{splncs04}
\bibliography{references}

\end{document}